\documentclass{aa}  

\DeclareRobustCommand{\VAN}[3]{#2}
\let\VANthebibliography\thebibliography
\def\thebibliography{\DeclareRobustCommand{\VAN}[3]{##3}\VANthebibliography}

\usepackage{graphicx}
\usepackage[switch]{lineno}
\usepackage{amsmath}
\usepackage{txfonts}
\usepackage{appendix}
\usepackage{hyperref}
\usepackage{xspace}
\hypersetup{colorlinks=true, citecolor=blue, linkcolor=blue}

\newcommand{\spaceAA}{\AA\xspace}
\newcommand{\msfd}{$\Sigma_*$\xspace}
\newcommand{\rsfr}{$\Sigma_{SFR}$\xspace}
\newcommand{\HII}{\hbox{H\,{\sc ii}}}
\newcommand{\ha}{\hbox{H$\alpha$}\xspace}
\newcommand{\hb}{\hbox{H$\beta$}\xspace}
\newcommand{\oii}{\hbox{[O\,{\sc ii}]}\xspace}

\newcommand{\nii}{\hbox{[N\,{\sc ii}]}\xspace}
\newcommand{\sii}{\hbox{[S\,{\sc ii}]}\xspace}

\newcommand{\Halpha}{\hbox{H$\alpha$} $\lambda$ 6563\xspace}

\newcommand{\NII}{\hbox{[N\,{\sc ii}]}$\lambda$ 6584\xspace}
\newcommand{\SIId}{\hbox{[S\,{\sc ii}]}$\lambda \lambda$ 6716,6731\xspace}
\newcommand{\OIId}{\hbox{[O\,{\sc ii}]}$\lambda \lambda$ 3726,3729\xspace}
\newcommand{\OIII}{\hbox{[O\,{\sc iii}]}$\lambda$ 5007\xspace}
\newcommand{\nsh}{N2S2H$\alpha$\xspace}

\newcommand{\rmunit}{$\log (\Sigma_*/( {\rm M_\odot kpc^{-2} }))$ \xspace}

\defcitealias{ji2022correlation}{JY22}

\begin{document} 

   \title{Re-evaluating the resolved mass-metallicity relation with a self-consistent metallicity calibration}
   \titlerunning{Self-Consistent rMZR}


   \author{Ziming Peng
          \inst{1}\fnmsep\thanks{zmpeng@link.cuhk.edu.hk}
          \and Renbin Yan \inst{1,2}\fnmsep\thanks{rbyan@cuhk.edu.hk}
          \and Zesen Lin \inst{3,1,2}
          \and Xihan Ji \inst{4,5}
          }

   \institute{
        Department of Physics, The Chinese University of Hong Kong, Shatin, New Territories, Hong Kong SAR, China\\
        \and
        CUHK Shenzhen Research institute, No.10, 2nd Yuexing Road, Nanshan, Shenzhen, China\\
        \and 
        Institute for Astrophysics, School of Physics, Zhengzhou University, Zhengzhou, 450001, China\\
        \and
        Kavli Institute for Cosmology, University of Cambridge, Madingley Road, Cambridge CB3 0HA, UK\\
        \and
        Cavendish Laboratory, University of Cambridge, 19 JJ Thomson Avenue, Cambridge CB3 0HE, UK
             }

   \date{Received XXX; accepted YYY}

  \abstract
   {}
   {The mass-metallicity relation (MZR) is essential for understanding the chemical evolution of galaxies. Whether the star formation rate (SFR) plays a role in setting the metallicity has long been debated. Using various metallicity calibrations can result in different conclusions for this fundamental yet unresolved issue. }
   {We apply a self-consistent metallicity calibration based on photoionization models to re-evaluate the resolved and integrated MZR. We utilize the integral field unit data from SDSS-IV/MaNGA, with $\sim 3.5\times10^6$ spaxels and $\sim$ 4550 galaxies. We compare our preferred metallicity calibration with several strong-line calibrations in the literature and direct method metallicity. We analyze the metallicity residual of MZR to evaluate the effects of SFR and apply the partial correlation coefficient to quantify the effects.}
   {The metallicity calibration we used shows the best consistency with the direct method. We provide 3 equations for resolved MZR, and verify that local SFR does not show significant correlation with metallicity. Considering the integrated properties, (s)SFR do not present correlation with the metallicity residuals. The results suggest that an equilibrium of inflow and outflow is favored, and the mass-metallicity relation does not have a secondary dependence on SFR. }
   {}


   \maketitle
%

\section{Introduction}

Knowing the gas-phase metallicity of galaxies is crucial for understanding their baryon cycle, hence the evolution of galaxies \citep{lilly2013gas,maiolino2019re}. In galaxies, gas-phase metallicity is determined simultaneously by several processes, including the stellar yields, gas inflow, and metal-rich gas outflow. The interplay of these processes leads to fundamental scaling relations, with the mass-metallicity relation (MZR) between stellar mass and gas-phase metallicity being one of the most explored and important relations.

The relation between stellar mass and metallicity was discovered first for irregular dwarf galaxies \citep{1979A&A....80..155L,1981ApJ...243..127K}. With the launch of Sloan Digital Sky Survey (SDSS, \citealt{2000AJ....120.1579Y}), \cite{2004ApJ...613..898T} presented a convincing mass-metallicity relation using a very large statistical sample. This relation has been extensively studied in both local galaxies (e.g., \citealt{kewley2008metallicity,andrews2013mass}) and galaxies in various redshift ranges (e.g., \citealt{2021ApJ...914...19S,2023ApJ...955L..18L,2023ApJ...957...39L,2024A&A...684A..75C,2025arXiv250805335A,2025A&A...699A...6P,2025AJ....170..307J}). It also extends to low-metallicity dwarf galaxies (e.g., \citealt{2017A&A...601A..95C,2023A&A...679A..98B,2024ApJ...976L..15C}). Meanwhile, various hydrodynamic simulations provides predictions of MZR \citep{2016MNRAS.456.2140M,2019MNRAS.484.5587T,2024ApJ...967L..41M,2025arXiv250118687Q}. With the development of integral field spectroscopy (IFS) in recent years, surveys like CALIFA \citep{sanchez2012califa}, SAMI \citep{bryant2015sami}, and MaNGA \citep{bundy2014overview} provide spatially resolved information of nearby galaxies, enabling studies of resolved stellar mass surface density (\msfd)-metallicity relation (rMZR, \citealt{2012ApJ...756L..31R,sanchez2012califa,2016MNRAS.463.2513B,2018ApJ...868...89G,2019MNRAS.484.3042S,2020ARA&A..58...99S,2022A&A...661A.112Y}). It is widely recognized that the metallicity increases with the increase of stellar mass, and this relation flattens at high stellar mass end (M$_*$>11 log($M_\odot$), \msfd>8 log($M_\odot$)/kpc$^2$), where the metallicity stops increasing, both globally and locally.

Various physical properties are introduced to explain the scatter of MZR. The most frequently discussed property is the star formation rate (SFR), or specific star formation rate (sSFR), defined as the SFR per unit stellar mass. Star formation can be enhanced by the metal-poor infalling gas, and it can drive metal-loaded outflow via stellar feedback, hence reducing local metallicity.  
For global MZR, galaxies that have higher SFRs (or sSFRs) tend to have lower metallicity at fixed stellar mass \citep{2008ApJ...672L.107E,2014ApJ...797..126S,2015ApJ...799..138S}. A fundamental plane minimizing the scatter of observed galaxies in the M-Z-SFR 3D space is constructed, and their relation is named the fundamental metallicity relation (FMR, \citealt{2010MNRAS.408.2115M,2010A&A...521L..53L}). A certain angle is found to minimize the metallicity scatters, which is slightly different in different studies, possibly due to different methods and mathematical models used \citep{2010A&A...521L..53L,2021MNRAS.506.1237T}. For resolved studies, some of them find a similar FMR as the global one \citep{2017MNRAS.468.4494Z,2025MNRAS.540.2667B}, while others find a positive or no correlation between SFR and metallicity \citep{2017ApJ...844...80B,2021ApJ...910..137W,2025arXiv250918961K}, or a more complicated situation \citep{2019ApJ...882....9S}. Moreover, some studies find that the resolved relation not only depends on local properties (\msfd, \rsfr, etc.), but also correlates with global properties (integrated stellar mass, total SFR, galaxy morphologies, etc., \citealt{2023MNRAS.519.1149B}). How these global and local properties affect the rMZR is still debated \citep{2019ApJ...878L...6S}.

The crux of getting different SFR dependence on rMZR is the use of varying metallicity calibrations (e.g., \citealt{2020ApJ...897...61T}). The direct method using electron temperature to derive metallicity is believed to be one of the most robust methods, but the faintness of auroral lines limits the scope of such studies in observations. While the direct method can still be applied with stacking of statistical samples \citep{andrews2013mass,2020MNRAS.491..944C}, spatial information is lost when the whole galaxy is integrated \citep{khoram2025direct}. To substitute for the direct method, empirical strong-line methods, which often use combinations of strong forbidden lines and hydrogen recombination lines, are used to measure metallicity \citep{2004MNRAS.348L..59P,dopita2013new,pilyugin2016new}. The relations between strong-line ratios and metallicities are calibrated based on galaxies with direct measurements. As a result, the empirical strong-line methods are not calibrated outside certain metallicity ranges, for example, at high metallicity (12+log(O/H) > 8.7, \citealt{easeman2024optimal}). Also, strong-line ratios might depend on other physical properties, such as the ionization parameter, diffuse ionized gas contamination \citep{zhang2017sdss}, and the chemical abundance pattern \citep{kewley2019understanding,2025A&A...699A.380C}, which makes it more difficult to ensure consistency across all calibrators. An alternative method is to use photoionization models to predict strong-line ratios for a range of metallicities. By matching the observed line ratios of a region to these model predictions, the corresponding input metallicity of the model can be assigned to that region \citep{mcgaugh1991h,kewley2002using}, but there are also limitations due to unreliable model assumptions, such as the ionizing spectra, prescription for secondary nitrogen abundances, and prescription for dust depletion. 

The limitations of the photoionization model can lead to misjudgments of the physical properties and their correlations inferred from the observed data. First, in the models, different combinations of metallicity and ionization parameter may yield the same value in some line ratios. This degeneracy will make the model appear folded and twisted when plotted in a line-ratio space, and the observed data will yield more than one set of derived physical properties, leading to errors. Second, when we derive metallicity from a set of models using observed data, we may get different results for different combinations of emission-line ratios used.
These situations emphasize the need for self-consistency across multiple line ratios --- that a given spectrum should correspond to the same model combination of metallicity and ionization parameter no matter which subset of line ratios we examine.  Traditional BPT diagrams can only visualize the consistency between data and model in two line ratios at a time, hence have difficulty judging such self-consistency across different BPT diagrams. \cite{ji2020constraining} provides a visualization method to judge whether a photoionization model can match the data in three line ratios simultaneously. They also provide a set of MaNGA-calibrated photoionization models that are self-consistent in multi-dimensional line-ratio space. 

In this work, we present a new study of the rMZR with a statistical sample from MaNGA using the self-consistent metallicity calibration method. \cite{ji2022correlation} (hereafter \citetalias{ji2022correlation}) provides metallicity measurements from the photoionization model mentioned above. Applying this photoionization model, one can derive consistent metallicity with different combinations of strong-line ratios. In addition, the \citetalias{ji2022correlation} metallicities are quite consistent with the results based on the direct method using MaNGA \citep{peng2025dmd}, with a much smaller discrepancy than previous works. These make \citetalias{ji2022correlation} metallicity a reliable choice for measuring rMZR. With the kpc-scale spatially resolved data covering a large number of galaxies from MaNGA, we are able to explore the rMZR and its dependence on other physical properties.

The paper is structured as follows. Section \ref{sec:2} introduces the data selection and the photoionization model. Section \ref{sec:3} describes how we derive the physical quantities shown in this paper, and compares different metallicity calibrations. Section \ref{sec:4} presents the rMZR result and the SFR dependence, both locally and globally. A new empirical calibration based on properly selected line ratios is presented in Section \ref{sec:5}. We conclude in Section \ref{sec:6}. 

For the brevity of writing, we use the abbreviations of line ratios in this paper. Unless otherwise specified, N2, S2, and R3 represent log(\NII/\ha), log(\SIId/\ha), and log(\OIII/\hb) in this paper, respectively. Throughout this paper, we assume the cosmology parameters of $\rm H_0= 73\,km~s^{-1}~Mpc^{-1}$, $\rm \Omega_m=0.3$, and $\rm \Omega_\Lambda=0.7$. A \cite{kroupa2001variation} Initial Mass Function (IMF) is assumed unless otherwise specified.


\section{Data}
\label{sec:2}

\subsection{MaNGA}
Our samples come from the SDSS Data Release 17 \citep{abdurro2022seventeenth}, which is also the 11th product launch of the Mapping Nearby Galaxies at Apache Point Observatory (MaNGA) survey \citep{bundy2014overview,yan2016sdss}. As one of the three main surveys of SDSS-IV \citep{2017AJ....154...28B}, MaNGA is an IFS survey that includes more than 10,000 galaxies in the local Universe, providing spatially resolved data for each of them. MaNGA's sample includes a primary sample observed out to a major axis radius of 1.5$R_e$, and a secondary sample extending to 2.5$R_e$ \citep{wake2017sdss}. MaNGA has a field of view (FoV) whose diameters vary from 12'' to 32'' \citep{drory2015manga}. MaNGA utilizes the BOSS spectrographs \citep{smee2013multi}, which have a wavelength range from 3622 to 10,354 \spaceAA and a spectral resolution of R $\sim$ 2000, covering optical and near-infrared information. The spatial resolution of the survey is $\sim$ 1-2 kpc. Observed data first go through the Data Reduction Pipeline (DRP), described by \cite{law2016data} and \cite{yan2016asdss}. After that, the extracted data cube are input into the Data Analysis Pipeline (DAP, \citealt{westfall2019data}) to fit stellar continuum using the Penalized PiXel-Fitting software (pPXF, \citealt{cappellari2017improving}) and to generate emission line fluxes \citep{belfiore2019data}. In this work, we use emission line flux measurements and equivalent width (EW) measurements from the DAP product.

\begin{figure*}
   \centering
	\includegraphics[width=\textwidth]{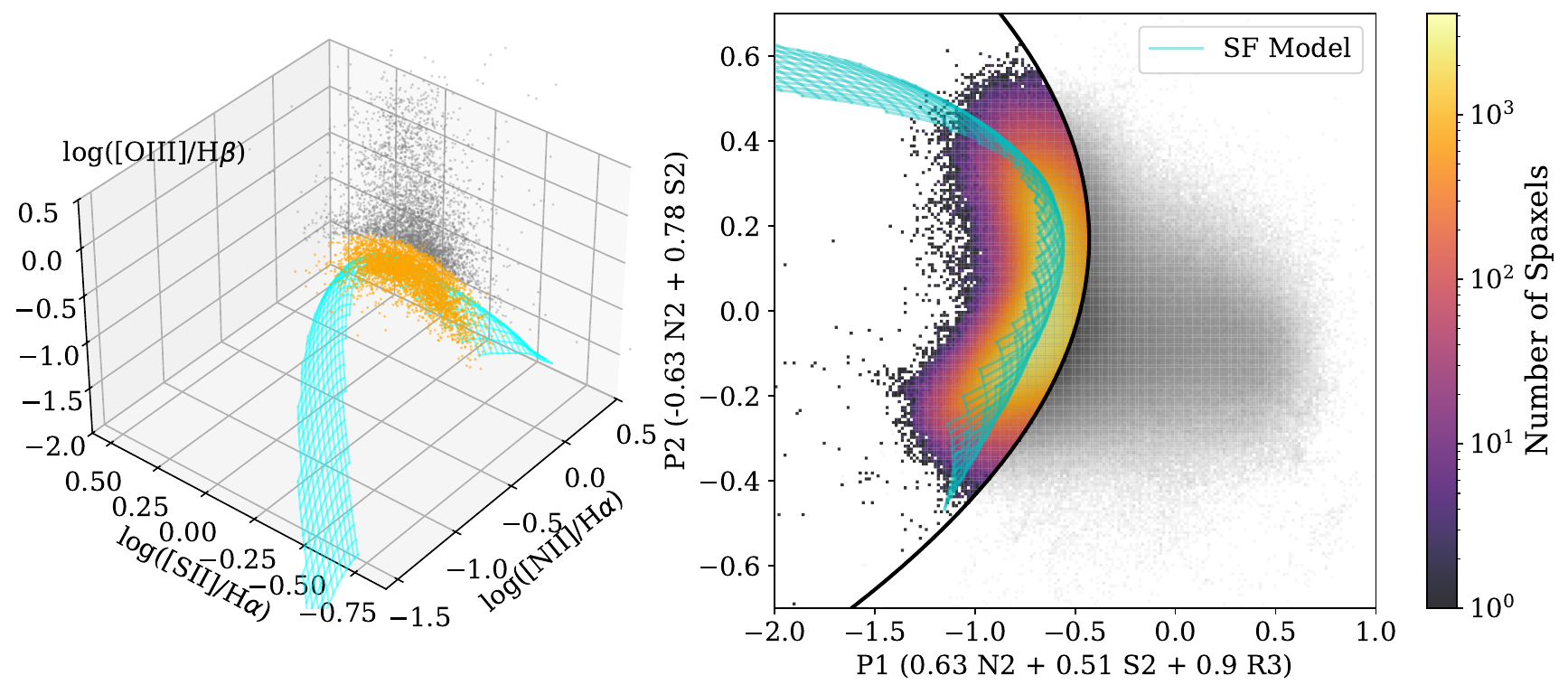}
    \caption{Left: An illustration of the 3D self-consistent photoionization models. The cyan grids show the models for star-forming regions; each node corresponds to a unique pair of metallicity and ionization parameter. The orange data points represent spaxels that are classified as star-forming regions, and the gray data points represent other spaxels. Right: Density distribution of the data (gray) and the star-forming spaxels (colored) in the refined optical diagnostic (P$_1$--P$_2$) diagram. The definitions of P$_1$ and P$_2$ are in the text. Cuts have been applied to the model grid so that only the parts that are within the middle 98\% of the data distribution along the hidden P$_3$ axis (perpendicular to P1 and P2) are shown. The black solid line represents the demarcation line. See \cite{ji2020constraining} for details. }
    \label{fig:2_3}
\end{figure*}

\subsection{Photoionization model and JY22 metallicity measurement}
\label{sec:2_2}

The photoionization models for \HII\ regions employed in this study were generated using version 17.03 of the \textsc{Cloudy} photoionization code \citep{ferland20172017}. These simulations assume an isobaric \HII\ region with a plane-parallel geometry. The ionizing spectral energy distribution (SED) was produced with Starburst99 v7.01 \citep{leitherer1999starburst99}. This SED corresponds to a stellar population with a continuous star-formation history (SFH) of 4 Myr, a \cite{kroupa2001variation} IMF computed with the \cite{pauldrach2001radiation} and \cite{hillier1998treatment} stellar atmospheres, and a standard Geneva evolutionary track. The hydrogen density was set to a constant of 14 cm$^{-3}$, a value derived from the median \sii $\lambda$ 6716 / \sii $\lambda$ 6731 ratio observed in \HII\ regions within the MaNGA survey \citep{ji2020constraining}. The stellar metallicity ranges available in Starburst99 are 0.05, 0.2, 0.4, 1.0, and 2.0 in $\rm Z/Z_\odot$. The gas-phase metallicity was set to match the stellar metallicity in each model and extended to approximately 3.16 (equivalent to 0.5 in logarithmic space) times the solar value, where the ionizing SED of the twice solar metallicity model is used. In these models, the abundances of heavy elements, except for helium (He), carbon (C), and nitrogen (N), are assumed to scale with oxygen according to the solar abundance pattern \citep{2010Ap&SS.328..179G}. The abundance of He follows the equation in \cite{2002ApJ...572..753D}. The abundances of the secondary elements C and N were calculated based on the N/O versus O/H relationship given by \cite{dopita2013new}, with the C abundance fixed at 1.03 dex above the N abundance before considering dust depletion. Dust depletion was subsequently incorporated using the default depletion factors provided in \textsc{Cloudy} \citep{cowie1986high,jenkins1987element}. The ionization parameter quantifies the ratios between ionizing radiation and density. It is represented by $\rm U = \frac{\phi_0}{n_H c}$, where $\phi_0$ is the ionizing flux, and $n_H$ is the volume density of hydrogen. In these photoionization models, we apply a grid of ionization parameter of -4.0 < log(U) < -0.5. Detailed descriptions of these photoionization models are available in \cite{ji2020constraining}.

For all the MaNGA star-forming pixels with spectra (hereafter spaxels), we obtain their N2, S2, and R3 ratios. We interpolate the photoionization model grid onto a 200 $\times$ 200 grid in the metallicity--ionization parameter plane. Given the model-predicted line-ratio variations across these two variables, and a data point in the line-ratio space with uncertainties, we can use the Bayesian method \citep{blanc2015izi} to derive the posterior distributions of metallicity and ionization parameter, assuming flat priors for the model parameters. A spaxel with ratios close to those of a model with a given metallicity and ionization parameter combination will have a high probability of having that metallicity and ionization parameter. We derive the metallicities and ionization parameters of the data from their posterior distribution-weighted mean values. In \citetalias{ji2022correlation}, N2O2 and O3O2 are also applied to derive the PDFs, yielding roughly consistent results as derived from N2, S2, and R3, which demonstrates the model also matches the \oii\ lines in the data simultaneously. We test different line-ratio combinations and find no significant differences in the results of this paper. Thus, we keep using N2, S2, and R3 to derive metallicities. 

We acknowledge that there are some fundamental physical assumptions in the model, such as using a specific ionizing SED, the dust depletion predictions \citep{2022MNRAS.512.2310G,2023MNRAS.520.4345G}, and the relation between secondary elements and metallicity. It is entirely possible that another model with different input assumptions may also fit these line ratios simultaneously. These limitations may indeed affect the final metallicity obtained. However, we believe that these can only have a minor effect on metallicity, since we use the joint constraint provided by 3 sets of strong line ratios. We verify the metallicity in Section \ref{sec:3} via the comparison with other metallicity calibrations. 

One may be concerned whether the integrated line intensity from \textsc{Cloudy} can be compared with spatially-resolved data. It is worth noting that the spatial resolution of MaNGA is $\sim$ 1 kpc, which is much larger than individual \HII\ regions. This allows us to compare the integrated line intensity of the \textsc{Cloudy} data with the spatially resolved data of MaNGA in this paper.

\begin{figure*}
   \centering
	\includegraphics[width=\textwidth]{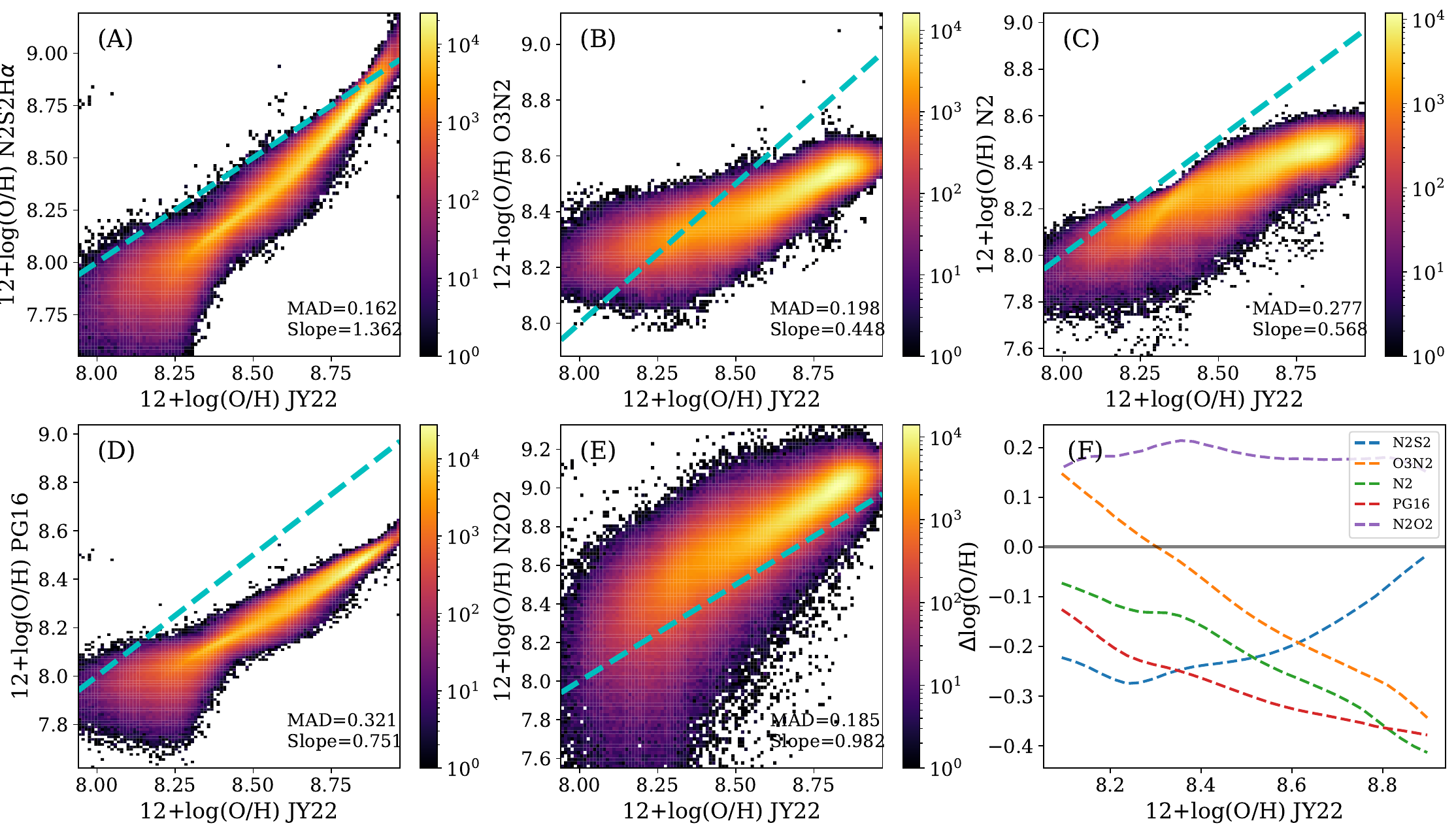}
    \caption{Comparisons between different metallicity indicators. Panel (A) to (E) show the \citetalias{ji2022correlation} metallicity (x axis) versus the empirical-calibrated strong-line metallicity  (A) \nsh \citep{dopita2016chemical}, (B) O3N2 \citep{marino2013o3n2}, (C) N2 \citep{marino2013o3n2}, (D) S-calibration (PG16, \citealt{pilyugin2016new}), and (E) N2O2 \citep{dopita2013new}.  Data are color-coded by the number densities of spaxels. The cyan dashed lines represent the 1:1 relation between the two axes. We annotate the median absolute deviation (MAD) and the best-fit slope between the two calibrations on the lower right of each plot. Panel (F) shows the median offsets between each empirical calibration and \citetalias{ji2022correlation} versus \citetalias{ji2022correlation} metallicity. Most of the calibrations show lower metallicities than \citetalias{ji2022correlation} and exhibit non-unitary slopes across the metallicity range. }
    \label{fig:3_2_1}
\end{figure*}

\begin{figure*}
   \centering
	\includegraphics[width=\textwidth]{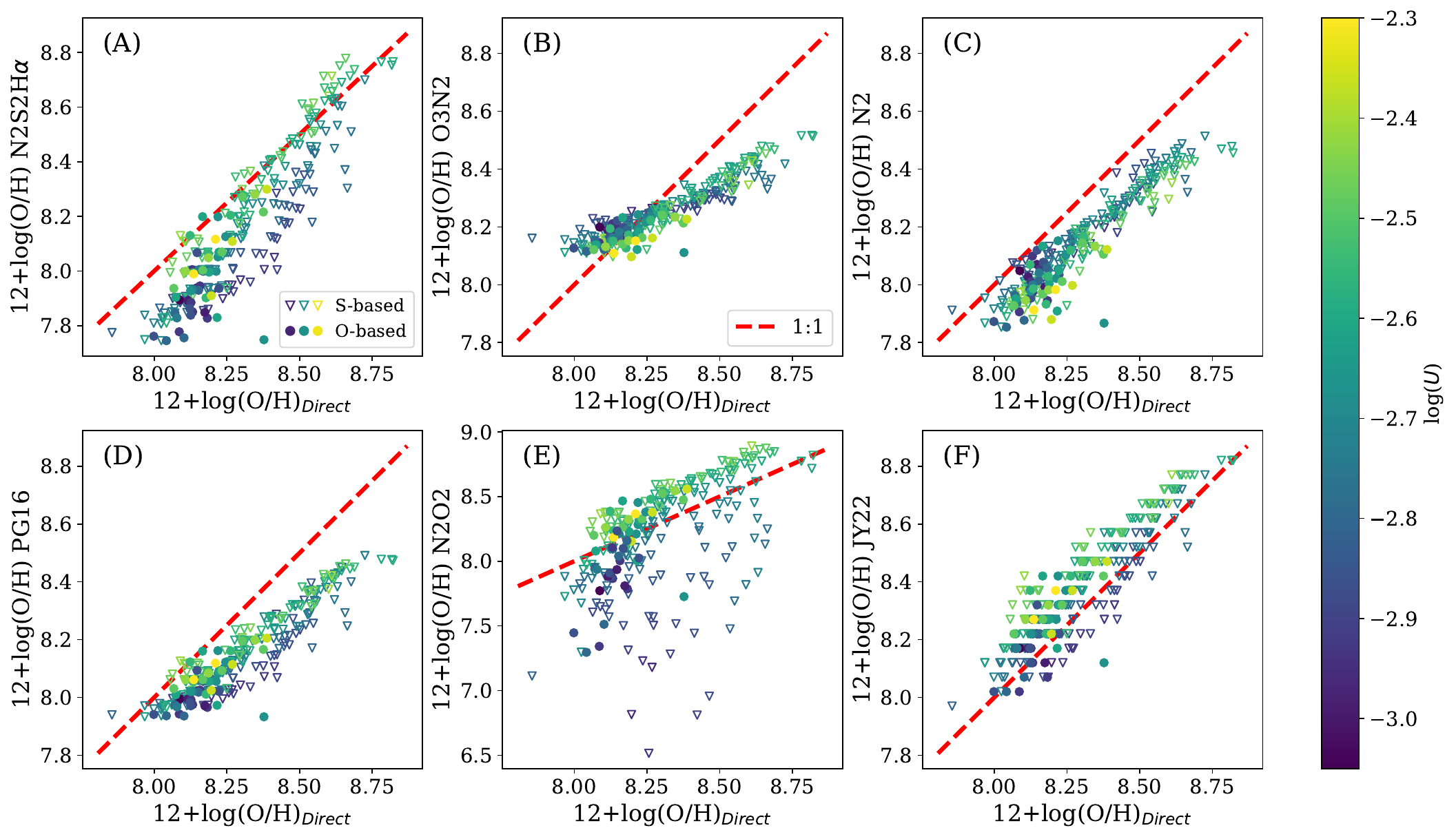}
    \caption{Comparisons between direct metallicity and different strong-line metallicity measurements. 
    Panels (A) to (E) use the same strong-line calibrations as Fig. \ref{fig:3_2_1}. Panel (F) shows the comparison between \citetalias{ji2022correlation} metallicity and the stacked direct-method result from \cite{peng2025dmd}. Dots are color-coded by the ionization parameter of each stacking bin. Inverted triangles represent the metallicities derived using oxygen electron temperatures, and circles represent the metallicities derived using sulfur electron temperatures. The dashed red lines represent equal values between the two axes. \citetalias{ji2022correlation} calibration shows the best consistency with the direct metallicities.}
    \label{fig:3_2_2}
\end{figure*}

\subsection{Sample selection}
\label{sec:2_3}

Since the \citetalias{ji2022correlation} metallicity focuses only on star-forming regions, we remove the contamination of active galactic nuclei (AGN) and low ionization emission regions (LIER) using a 3-dimensional diagnostic diagram from \cite{ji2020constraining}. This diagram is a combination of traditional Baldwin-Philips-Terlevich (BPT) \nii and \sii diagrams \citep{baldwin1981classification,veilleux1987spectral}, with N2, S2, and O3 values as the 3 axes. This 3D diagnostic diagram can avoid ambiguous classifications, i.e., contradicting classification resulting from different BPT diagrams (e.g., \citealt{vogt2014galaxy}). A new projection angle is chosen to make the photoionization model grids for both the star-forming regions and AGNs appear nearly edge-on, to make it easier to judge the data-model consistency in three line ratios simultaneously, and to separate star-forming regions and AGNs more cleanly. The left panel of Fig. \ref{fig:2_3} presents the distribution of spaxels in the 3D line ratio space, and the locations of the star-forming model grids. The photoionization model grid, as viewed edge-on, goes through the densest area of spaxels, meaning that for most data points on the star-forming locus, one can always find a model grid point that can simultaneously match all three line ratios. The right panel of Fig. \ref{fig:2_3} plots a projection of the 3D space in a certain viewing angle (a polar angle $\theta=36^{\circ}$ from the R3 axis and a counter-clockwise azimuthal angle $\phi=39^\circ$ from the N2 axis). The P$_1$ and P$_2$ are defined as

\begin{equation}
   P_1 = 0.63\,\rm N2+0.51\,S2+0.59\,R3
	\label{eq:p1}
\end{equation}
and
\begin{equation}
   P_2 = -0.63\,\rm N2+0.78\,S2\, .
	\label{eq:p2}
\end{equation}

The demarcation line corresponds to $f_{SF}=0.90$, which means the maximum contamination to \ha from AGN-ionized regions is 10\%.
A line is adopted to select star-forming regions and is given by

\begin{equation}
   P_1<-1.57P_2^2 + 0.53P_2 - 0.48.
	\label{eq:sf}
\end{equation}

We apply a signal-to-noise ratio (S/N) cut to all the strong lines that are involved in measuring metallicity (\ha, \hb, \NII, \SIId, \OIII, and \OIId). Any spaxel with one of these emission lines having S/N < 3 is excluded. We also apply an EW cut for \ha to avoid the contamination from DIG-dominated spaxels. Spaxels with EW(\ha) < 6 \AA\ are excluded, following \cite{lacerda2018diffuse}. 
Finally, we obtain $\rm 3.42\times10^6$ spaxels from 5,058 galaxies.

\section{Deriving and comparing physical properties}
\label{sec:3}

\subsection{Deriving global and local properties}

The total stellar mass of each galaxy is retrieved from the MaNGA Pipe3D \citep{2022ApJS..262...36S} Value-Added Catalog (VAC). They are extracted from the NSA catalog\footnote{https://www.sdss4.org/dr17/manga/manga-target-selection/nsa/}. The stellar mass surface density (\msfd) of each spaxel is also from Pipe3D. The Pipe3D \msfd is calculated with the summed mass-to-light ratio weighted by different simple stellar populations (SSPs), assuming a \cite{1955ApJ...121..161S} IMF. It provides the measurement of \msfd per spaxel. Since we also know the angular diameter distance of galaxies, we can obtain the \msfd in the unit of $M_\odot$ per kpc$^2$ after converting the spaxel edge length (0.5 arcsecond) to physical scale and correcting for inclination following \cite{2016MNRAS.463.2513B}. 
The total SFR of each galaxy is also obtained from Pipe3D, which applied the dust-corrected \ha flux to calculate the SFR of each spaxel using the \cite{1998ARA&A..36..189K} relation, then integrated them. We applied the IMF conversion from \cite{1955ApJ...121..161S} to \cite{kroupa2001variation} for the total SFR provided by Pipe3D. 
The total sSFR of each galaxy is the total SFR divided by the total stellar mass. 
The SFR surface density (\rsfr) is derived from the \ha flux of each spaxel. Before computing \rsfr, we correct for the dust attenuation effect of galaxies assuming $\rm R_v=3.1$, using the empirical curve from \cite{fitzpatrick1999correcting}. Changing the empirical extinction curve to \cite{cardelli1989relationship} or \cite{calzetti2000dust} does not affect the results in this paper. 

We follow \cite{2012ARA&A..50..531K} to measure the \rsfr,

\begin{equation}
  \rm  log(SFR\ /M_{\odot} \,yr^{-1}) = log(L_{H\alpha}\ /erg\, s^{-1}) - 41.27,
\end{equation}
which is an IMF-converted version for \cite{1998ARA&A..36..189K}. The resolved sSFR is derived by 

\begin{equation}
  \rm  sSFR\ (yr^{-1})= \frac{\Sigma_{SFR}}{\Sigma_*}\,.
\end{equation}

In the figures shown in this paper, we use sSFR$_T$ to represent the total sSFR, and sSFR$_R$ to represent the spatially resolved sSFR. 

\subsection{Gas-phase metallicity}
\begin{table*}
   \caption{Five empirical strong-line calibrators used in this section}
   \label{table:3_2}      
   \centering          
   \begin{tabular}{l l c }     
   \hline\hline       
   Abbreviation &Reference& Equation\\ [0.3em]
      \hline                   
      N2S2H$\alpha$ &\cite{dopita2016chemical}& 12 + log(O/H) = 8.77 + N2S2 + 0.264 $\times$ N2H$\alpha$ \\ [1em]
      O3N2 &\cite{marino2013o3n2}& 12 + log(O/H) = 8.533 - 0.214 $\times$ O3N2 \\  [1em]
      N2 &\cite{marino2013o3n2}& 12 + log(O/H) = 8.743 + 0.462 $\times$ N2H$\alpha$ \\ [1em]
       & & 12 + log(O/H) = 8.424 + 0.030 $\times$ O3S2 + 0.751 $\times$ N2H$\beta$ \\
      PG16 (S-calibration)& \cite{pilyugin2016new} & + (-0.349 + 0.182 $\times$ O3S2 + 0.508 $\times$ N2H$\beta$) $\times$ S2H$\beta$  (high)\\ [0.5em]
       & & 12 + log(O/H) = 8.72 + 0.789 $\times$ O3S2 + 0.726 $\times$ N2H$\beta$ \\
       & & + (1.069 - 0.170 $\times$ O3S2 + 0.022 $\times$ N2H$\beta$) $\times$ S2H$\beta$  (low)\\ [1em]
      N2O2 &\cite{dopita2013new}& 12 + log(O/H) = log(1.54020 + 1.26602 $\times$ R + 0.167977 $\times$ R$^2$) + 8.93 \\
      & & (R = $\rm \frac{[N II]\lambda 6583}{[O II]\lambda \lambda 3726,3729}$)
      \\ [0.5em]
   \hline
   \end{tabular}
   \tablefoot{N2H$\alpha$ represents log(\NII/\ha). N2S2, O3N2, and O3S2 represent log(\NII/\SIId), log(\OIII/\NII), and log(\OIII/\SIId), respectively. N2H$\beta$ and S2H$\beta$ represent log(\NII/\hb) and log(\SIId/\hb). R in the N2O2 row means the linear ratio of \NII and \OIId. The high and low in PG16 stand for N2H$\beta$ $\geq$ -0.6 and N2H$\beta$ < -0.6.}
\end{table*}

The importance of selecting an appropriate metallicity calibrator cannot be overemphasized for obtaining the correct mass-metallicity relation. In this section, we select five strong-line calibrators with different emission-line ratios and compare them with the \citetalias{ji2022correlation} metallicity we will use. One of the most popular calibrations used in previous rMZR studies is the O3N2 method from \cite{marino2013o3n2}. We select both the O3N2 and N2 calibrations proposed by \cite{marino2013o3n2}, since they have better agreement with other diagnostics than the O3N2 and N2 calibrations that \cite{2004MNRAS.348L..59P} provides \citep{2021MNRAS.502.3357P}. We also select the \nsh calibration provided by \cite{dopita2016chemical}, which is considered the most accurate calibration by \cite{easeman2024optimal}. The S-calibration involving \NII, \OIII, \hb, and \SIId as described by \cite{pilyugin2016new}, and the N2O2 calibration from \cite{dopita2013new}, are frequently employed calibration methods in recent IFS surveys (e.g., \citealt{2022ApJ...929..118G,2023MNRAS.520.4902G,2025A&A...701A.226V}). Therefore, we also consider these two in our discussion. 

Table \ref{table:3_2} shows the line ratios used in each empirical calibration and the equations to calculate metallicities. For line ratios with small wavelength differences, such as N2H$\alpha$ and N2S2, the effect of dust attenuation is relatively negligible. Thus, we do not correct for the dust effect for them. For line ratios with large wavelength differences, such as N2H$\beta$, O3S2, and N2O2, we apply the \cite{fitzpatrick1999correcting} dust extinction correction, assuming a Case B recombination at 10,000 K, according to the Balmer decrement (\ha/\hb) of each spaxel \citep{2024A&A...691A.201L}.

Fig. \ref{fig:3_2_1} shows the correlations between the \citetalias{ji2022correlation} metallicities and the 5 strong-line metallicities. The dashed cyan line in each panel presents the 1:1 relation. When color-coded by their number densities, all 5 strong-line metallicities show significant deviations from the 1:1 relation with \citetalias{ji2022correlation} metallicity. Panel (F) of Fig. \ref{fig:3_2_1} demonstrates the median offsets between each empirical strong-line metallicity and \citetalias{ji2022correlation}, as a function of \citetalias{ji2022correlation} metallicity. One can see that N2O2 calibration has a roughly constant offset of $\sim$ 0.18 dex, while all other calibrations have non-unity slopes relative to \citetalias{ji2022correlation}. O3N2, N2, and PG16 metallicities all have slopes less than 1, meaning the metallicity ranges obtained using these calibrations will be more compressed and metallicity gradient shallower. On the other hand, \nsh has a slope steeper than 1 when compared to \citetalias{ji2022correlation}, meaning it will result in a more stretched metallicity difference and larger metallicity gradient. Among the 5 metallicity calibrations, \nsh has the smallest median absolute deviation (MAD), 0.161, and most of the data points lie slightly below the 1:1 relation. N2O2 also has a relatively small MAD, 0.180, and it is the only calibration whose metallicity is systematically higher than \citetalias{ji2022correlation}. O3N2, N2, and PG16 all yield lower metallicities than \citetalias{ji2022correlation}, while PG16 shows the largest MAD. \nsh and PG 16 show tighter correlations with \citetalias{ji2022correlation}, and the other three metallicity calibrations have slightly larger scatters. 
Fig. \ref{fig:3_2_2} presents the comparisons between strong-line methods and the direct method, using data from \cite{peng2025dmd}. The x-axis represents the direct metallicity measurement from stacked spectra, which are assembled from spaxels with similar N2, S2, and R3 ratios, assuming that similar line ratios correspond to similar intrinsic gas-phase metallicities. The dashed red line in each panel presents the 1:1 relation. Metallicities derived from oxygen electron temperatures are marked as circles, and metallicities derived from sulfur electron temperatures are marked as filled pluses. All the dots are color-coded by their \citetalias{ji2022correlation} ionization parameters. Panels (A) to (E) have the same y-axis as in Fig. \ref{fig:3_2_1}, while the y-axis of Panel (F) is the \citetalias{ji2022correlation} metallicity.

The \nsh method shows the most consistency with \citetalias{ji2022correlation}, as the data are distributed in a narrow range, and the relation is close to 1:1. \nsh metallicities are generally lower than \citetalias{ji2022correlation} metallicities, but they reach the same value when metallicity is high. It is also somewhat consistent with the direct method. In the low metallicity regime, \nsh predicts lower metallicity than the direct method. In contrast, these two methods almost give the same result in the metal-rich regime. 
The O3N2 presents a shallower increase than \citetalias{ji2022correlation} and the direct method. When metallicities are low, \citetalias{ji2022correlation} and O3N2 metallicities scatter around the 1:1 line. However, in the metal-rich regime (12+log(O/H) $\sim$ 8.9), O3N2 metallicities are significantly lower than \citetalias{ji2022correlation} by $\sim$ 0.3 dex. Similar to O3N2, N2 metallicities are also generally lower than \citetalias{ji2022correlation}, particularly in the high metallicity range. This is similar to what \cite{easeman2024optimal} reports. Interestingly, the densest regions in the \citetalias{ji2022correlation}-N2 plot are not linearly distributed, different from other calibrators. They exhibit a steep curve in 8.25 < 12+log(O/H)$_{JY22}$ < 8.5, where N2 metallicities increase from 7.95 to 8.3, then turn into a shallow curve at higher metallicities. Comparing these two calibrators to the direct method, O3N2 shows a similarly shallower slope and a complex secondary relation with the ionization parameter. N2 shows a negative secondary relation with ionization parameters, with those biased to lower metallicities having higher ionization parameters.
PG16 metallicities are distributed almost parallel with the 1:1 curve, with an offset of $\sim$ 0.25 dex from \citetalias{ji2022correlation} metallicities. This method shows the least scatter at high metallicities. However, the highest value from PG16 is $\sim$ 8.65, lower than the solar metallicity 8.69 \citep{asplund2009chemical}, meaning that all the spaxels we selected have sub-solar metallicities. It is significantly lower even than the direct method, which is always argued to underestimate the metallicity. 
For O3N2, N2, and PG16, the highest metallicities they reach is $\sim$ 8.65, meaning that there are $\sim$ 0.4 dex deviations from both \citetalias{ji2022correlation} and the direct method. 
N2O2 metallicities are generally higher than \citetalias{ji2022correlation} by $\sim$ 0.3 dex, and are also higher than the other four empirical calibrators. However, N2O2 also shows large scatter, especially in the low metallicity regime, where the difference between \citetalias{ji2022correlation} and N2O2 can reach a maximum of 0.5 dex. 

Compared to all five strong-line calibrators, the \citetalias{ji2022correlation} metallicity measurement shows the best consistency with the direct method. As demonstrated by \citetalias{ji2022correlation}, this photoionization model is the best among several models at simultaneously matching multiple strong-line ratios. Now we show that metallicity derived from this model also matches the one from the direct method quite well, meaning it could also match the auroral-to-strong-line ratios simultaneously. This is what we refer to as a self-consistent model because such models allow us to obtain consistent results no matter which subset of line ratios is used in deriving metallicity. We will apply \citetalias{ji2022correlation} to calculate metallicity and measure the scaling relations in the next section.

\begin{figure}
	\includegraphics[width=\columnwidth]{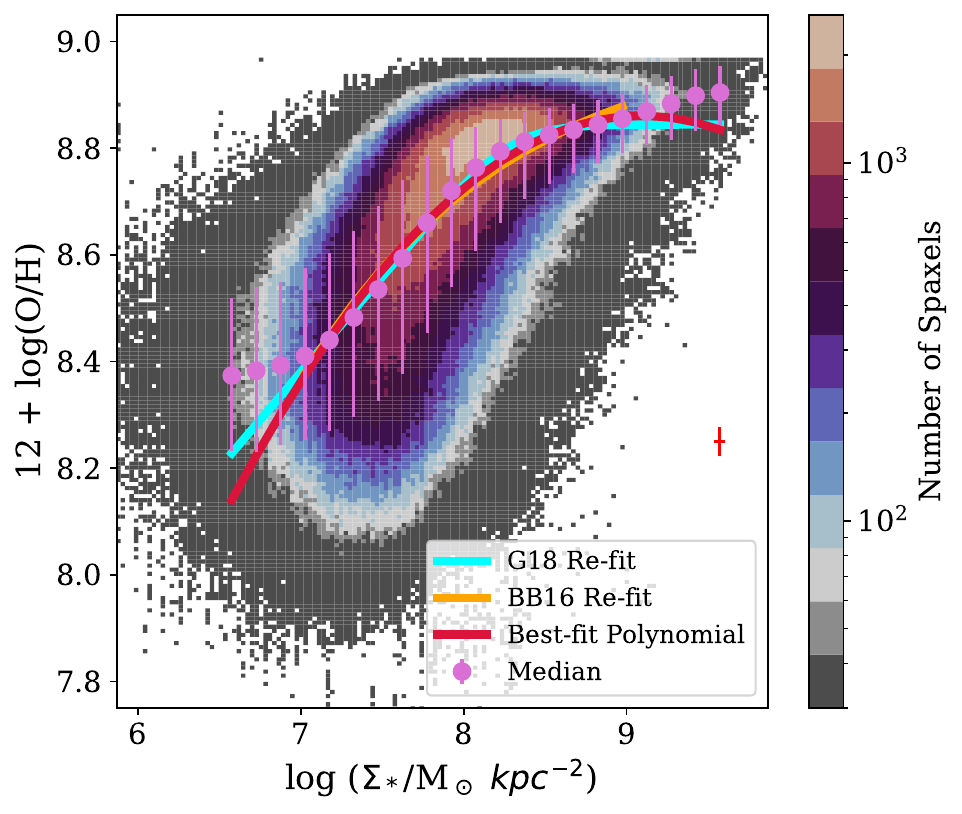}  
    \caption{Resolved mass-metallicity relation (rMZR) for MaNGA star-forming regions. The pink points with vertical bars show the median values and scatters of the rMZR (see text for detailed descriptions). The orange and cyan lines represent our re-fittings using the functional forms from BB16 \citep{2016MNRAS.463.2513B} and G18 \citep{2018ApJ...868...89G}, respectively. \textit{Note that these are significantly different from their original formula due to systematic differences in the metallicity calibrators.} The red cross on the lower right of the figure represents the median uncertainty of \msfd and metallicity.}
    \label{fig:4_1}
\end{figure}

\begin{table*}
   \caption{Three best-fit equations of rMZR.}
   \label{table:4_1}      
   \centering          
   \begin{tabular}{l c c c c c}     
   \hline\hline       

   Reference & Equation & a & b & c & d\\ [0.2em]
      \hline                   
      \cite{2016MNRAS.463.2513B} & $y=a+b(x-c)e^{-(x-c)}$ &9.01$\pm$0.044&-12.13$\pm$27.6&2.571$\pm$2.99&--\\[0.5em]
      \cite{2018ApJ...868...89G} & $y=a - b\times log(1 + (\frac{10^c}{10^x})^d)$ &8.88$\pm$0.012&0.116$\pm$0.048&8.31$\pm$0.12&1.43$\pm$0.44\\[0.5em]
      Polynomial & $y=a+bx+cx^2$ & 0.441$\pm$0.65 & 1.82$\pm$0.016 & -0.0979$\pm$0.0097  & --\\
   \hline
   \end{tabular}
\end{table*}

\begin{figure}
	\includegraphics[width=\linewidth]{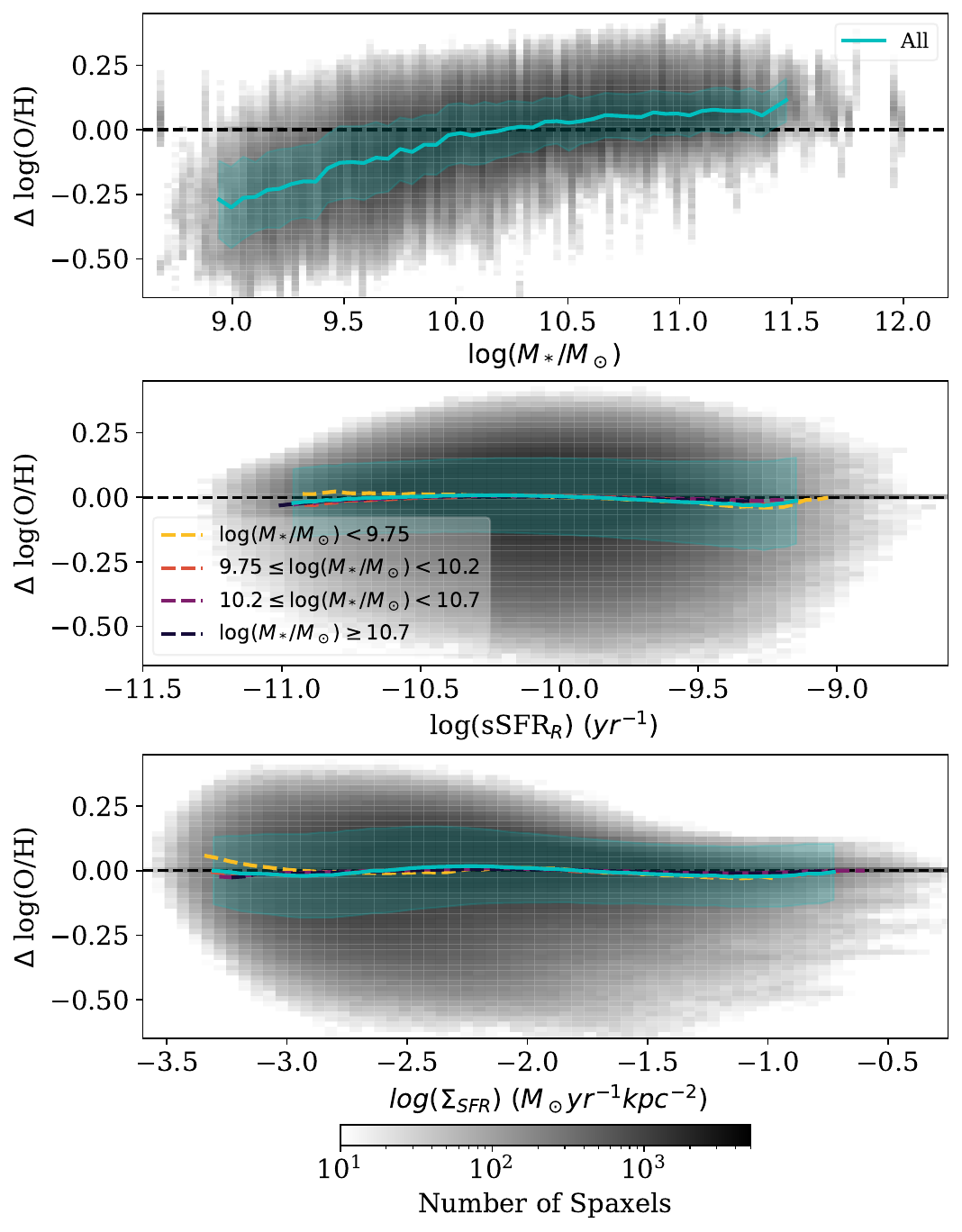}
    \caption{Scatter of the rMZR versus total mass (top panel), local sSFR (middle panel), and \rsfr (bottom panel). The 2D histograms show the distribution of the data. The solid cyan lines represent the median values of the data, and the cyan band indicates their standard deviation. For the middle and bottom panels, the dashed color lines represent the median values of the data in different total mass bins.}
    \label{fig:4_2_dmet}
\end{figure}

\begin{table}
   \caption{Partial correlation coefficients between metallicity residual and total mass, \rsfr, and sSFR, while controlling for the \msfd following the binning as given in Fig. \ref{fig:4_2_dmet}.}
   \label{table:4_2}      
   \centering          
   \begin{tabular*}{\linewidth}{l c c c}     
   \hline\hline       

   Mass Bin & M$_*$ & \rsfr & sSFR$_R$\\ [0.2em]
    \hline                   
      All & 0.69 & -0.01 & -0.04 \\[0.5em]
      log(M$_*$/M$_\odot$) < 9.75 & -- & -0.05 &  -0.05\\[0.5em]
      9.75 $\leq$ log(M$_*$/M$_\odot$) < 10.2 & -- & 0.03 & -0.01\\[0.5em]
      10.2 $\leq$ log(M$_*$/M$_\odot$) < 10.7 & -- & 0.01 & -0.01\\[0.5em]
      log(M$_*$/M$_\odot$) $\geq$ 10.7 & -- & 0.01 & -0.01\\
   \hline
   \end{tabular*}
   \tablefoot{For each $\rho_{ZX|Y}$ in the table, Y represents \msfd and Z represents the residual of metallicity. X is indicated in the first row. For those $\rho$ whose absolute values are less than 0.01, we retain the signs and write them as 0.01.}
\end{table}

\section{Results}
\label{sec:4}

\subsection{Resolved mass-metallicity relation}

Fig. \ref{fig:4_1} shows the spatially resolved mass-metallicity relation using the MaNGA star-forming spaxels and the \citetalias{ji2022correlation} metallicity calibration. The \rmunit ranges from 5.8  to 9.8, and the metallicity ranges from 7.75 to 8.95. We show the median metallicities at fixed \msfd with a range of 0.15 dex, and the error bars represent the 16th and 84th values of each \msfd bin. Previously, the rMZR was derived by \cite{2016MNRAS.463.2513B} (hereafter BB16) using part of the MaNGA galaxies, and the fitting equation is from \cite{2011arXiv1112.3300M}. \cite{2018ApJ...868...89G} (hereafter G18) also fitted rMZR using a different equation. BB16 has a \msfd range of 7 $\leq$ \rmunit $\leq$ 9 and saturates with a metallicity of 8.5 at the high \msfd end, while the rMZR of G18 has a rapid growth in metallicity at low \msfd, and saturates at a metallicity of $\sim$ 9.2. We pointed out in Sec. \ref{sec:3} that the difference is mainly made by using different metallicity calibrators \citep{kewley2008metallicity,2017MNRAS.469.2121S}. 
We applied the functional forms (shown in Table \ref{table:4_1}) from both BB16 and G18 (also used in \citealt{2020MNRAS.491..944C}) to fit the median data in Fig. \ref{fig:4_1}, which are significantly offset from the original formula given by those authors. Also, we applied the quadratic function fit used by \cite{2004ApJ...613..898T} and \cite{2019MNRAS.484.3042S}. To keep the consistency with previous works, we only selected 7.0 $\leq$ \rmunit $\leq$ 9.5  to fit the equations. Fitting results are shown in Table \ref{table:4_1}. They are also demonstrated on the solid orange (BB16), cyan (G18), and red (polynomial) in Fig. \ref{fig:4_1}.
As these three best-fit curves are highly similar and all close to the median curve, we use the median curve throughout this study to investigate the secondary dependence. Using any of these fittings would not change the conclusions in this study. 
We also notice that metallicity flattens at the low \msfd end. This may be due to the measurement uncertainty in \msfd, which is quite large, causing many intrinsically high \msfd points to be scattered into the low \msfd regime, artificially flattening the trend. Another potential cause is that the low-metallicity regions often come from outskirts of galaxies, where they tend to show less variation compared to the central region, resulting in a flatter relation \citep{2017MNRAS.469..151B}.

\begin{figure*}
    \centering
	\includegraphics[width=\textwidth]{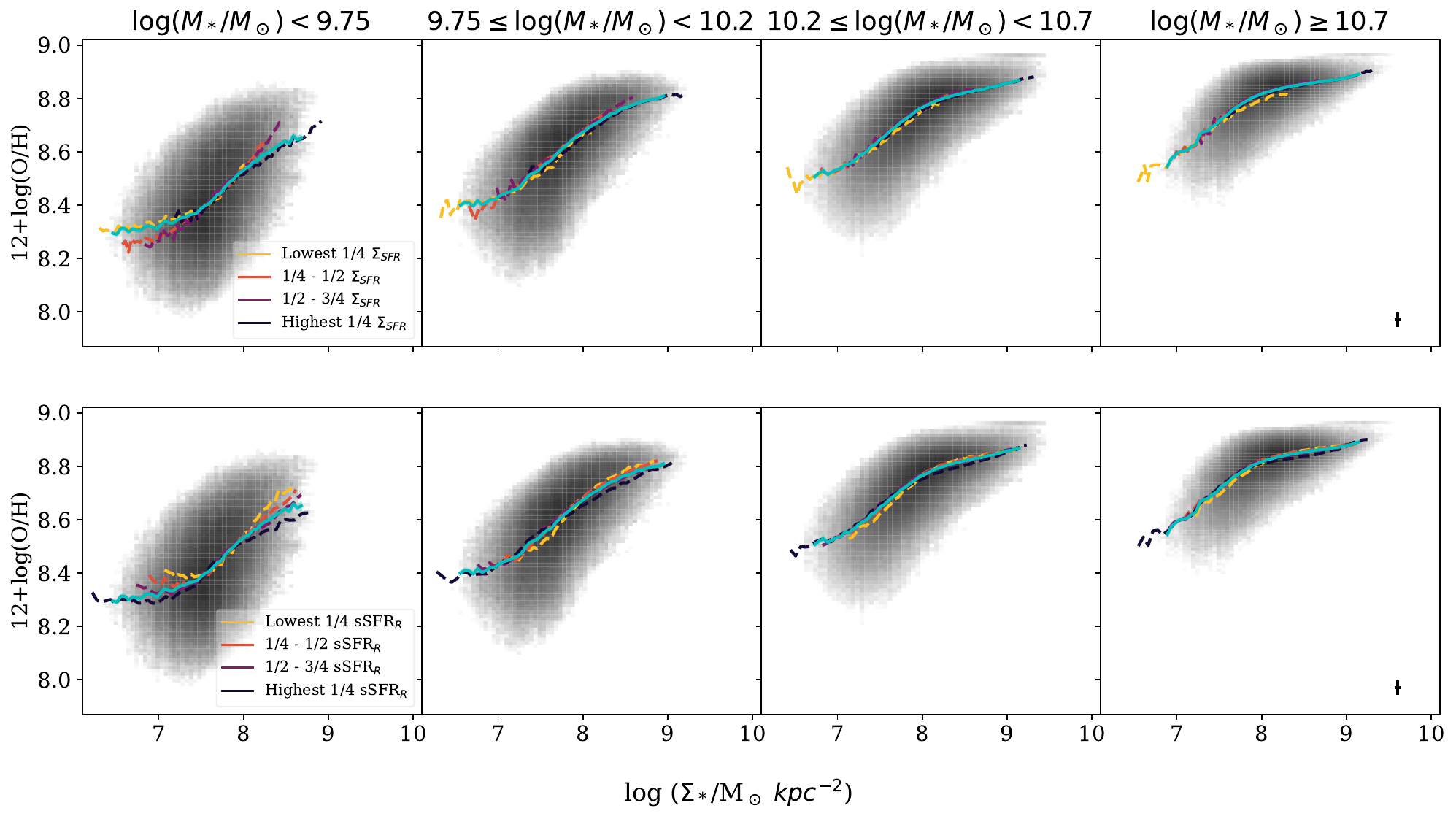}
    \caption{The rMZR binned by galaxy total stellar mass. From left to right, they are bins of $\log(M_*/M_\odot)<9.75$, $9.75\leq\log(M_*/M_\odot)<10.2$, $10.2\leq\log(M_*/M_\odot)<10.7$, and $\log(M_*/M_\odot)\geq10.7$. The cyan line in each panel demonstrates the median of all the data in this bin. The other four colored lines show the median of the data in a certain \rsfr bin. Except for the $\log(M_*/M_\odot)<9.75$ bin, different \rsfr shows almost no differences on rMZR.}
    \label{fig:4_2_3}
\end{figure*}

\subsection{Does the resolved fundamental metallicity relation exist?}

Given the large scatter of Fig. \ref{fig:4_1}, we explore the existence of any secondary dependence of metallicity. The most commonly discussed parameter is SFR, and it is still contentious whether a resolved FMR exists \citep{2020ApJ...897...61T,2024arXiv241204541N}. We also notice that a galaxy's total stellar mass plays a crucial role in determining its metallicity \citep{2018ApJ...868...89G,2023MNRAS.519.1149B}. A larger total mass for a galaxy means a heavier dark matter halo \citep{2018ARA&A..56..435W}, which enhances its ability to keep metals from outflow. Thus, they are likely to have higher metallicities. Also, total mass reflects the evolutionary stage of the galaxy. Galaxies with higher total mass tend to experience more mergers and other environmental effects \citep{2010ApJ...721..193P,2014ARA&A..52..291C}.

\begin{figure*}
    \centering
	\includegraphics[width=\textwidth]{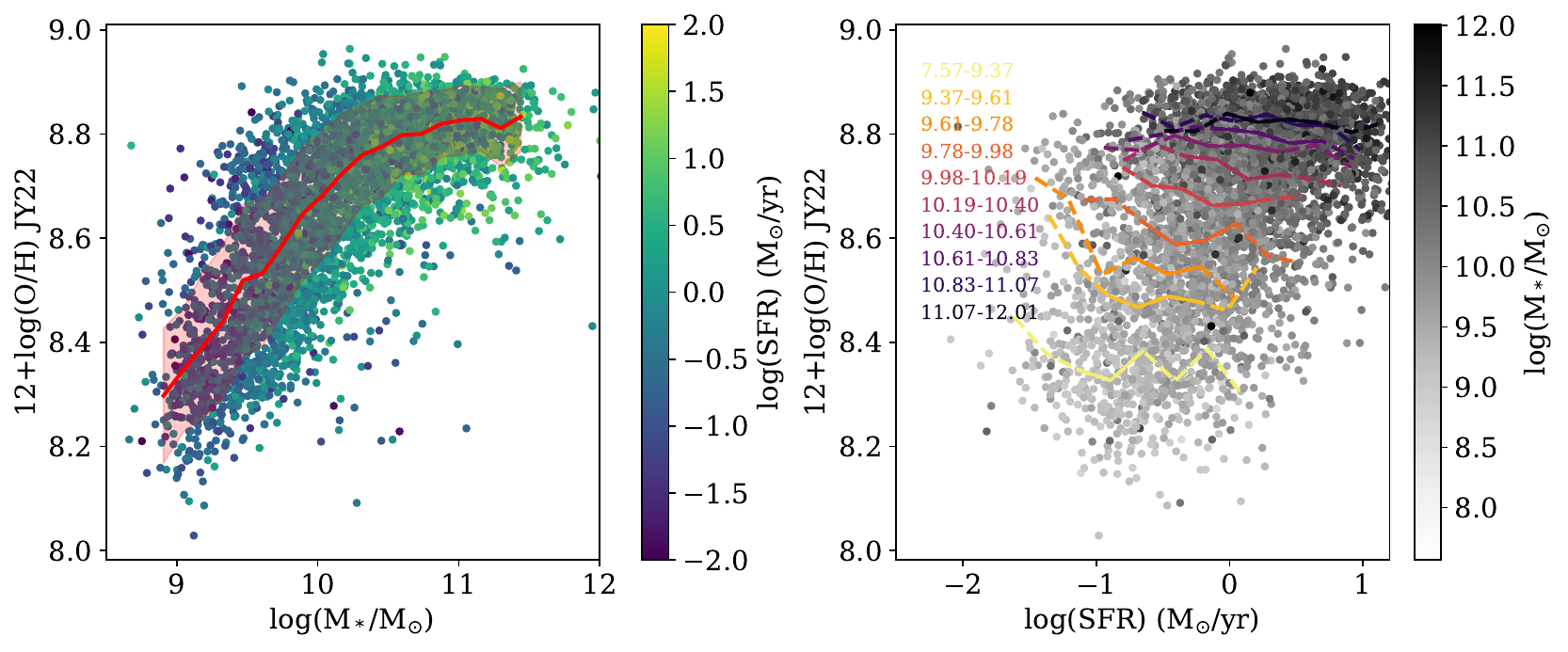}
    \caption{\textit{Left:} Integrated MZR for galaxies with more than 50 spaxels selected as star-forming spaxels (see text for details). Each dot stands for a galaxy. Dots are color-coded by the galaxy's total SFR. The red line represents the median values for different fixed total masses. \textit{Right:} Integrated SFR versus metallicity for the same galaxy samples. Dots are color-coded by the galaxy's total stellar mass. Color lines show the median metallicities as a function of SFR in each M$_*$ bin, with the M$_*$ range annotated in the figure using the corresponding color. Solid lines represent median values derived from over 50 data points, while dashed lines represent median values derived from fewer than 50 data points. Median values derived from fewer than 10 data points are not shown.}
    \label{fig:4_3_1}
\end{figure*}

\begin{figure*}
    \sidecaption
	\includegraphics[width=12cm]{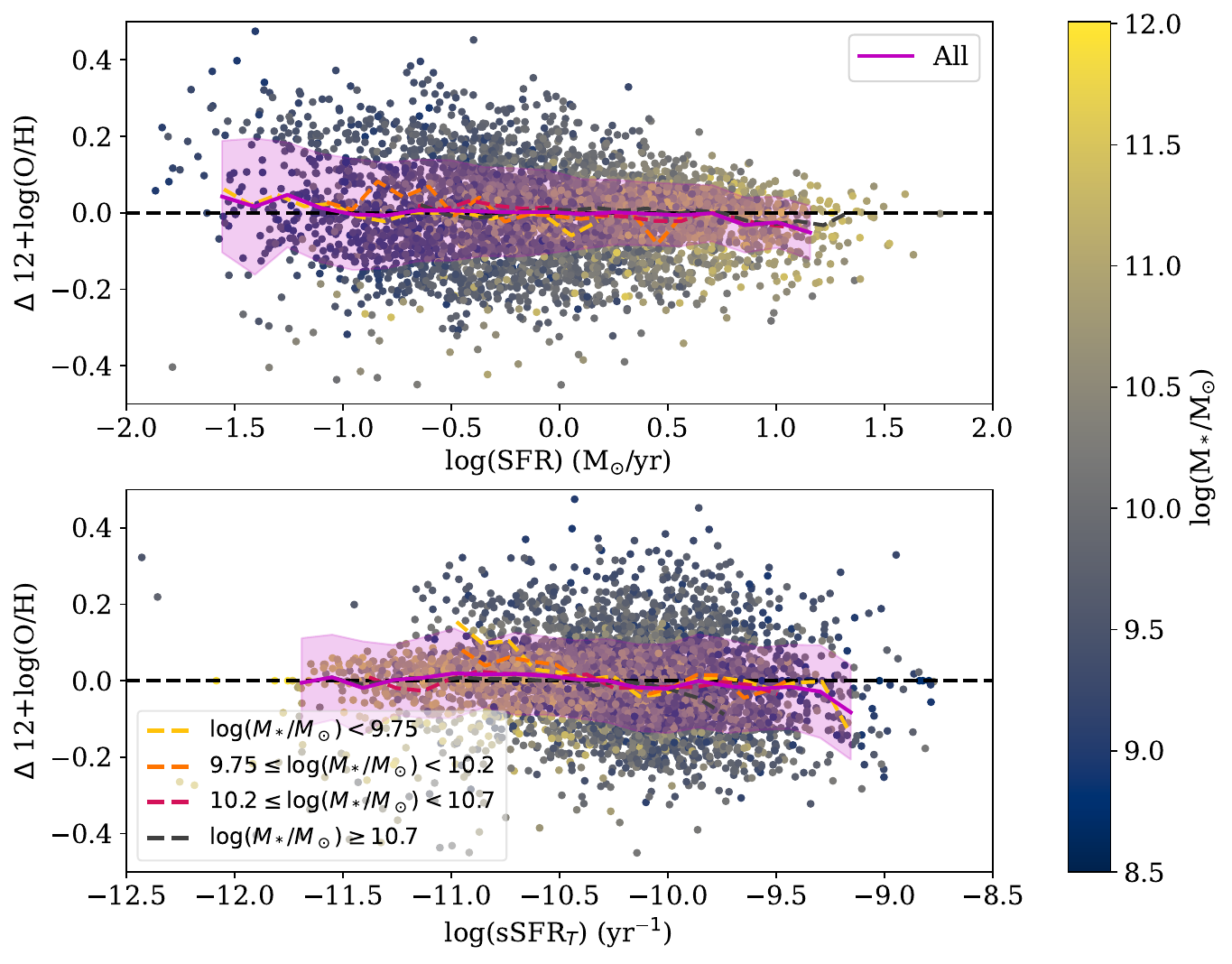}
    \caption{Integrated SFR versus the residual of integrated MZR (see left panel of Fig. \ref{fig:4_3_1}). Each dot represents one galaxy and is color-coded by its total stellar mass. The dashed black horizontal line indicates zero residual. The solid magenta line shows the median values of residual, with the shaded regions showing the 16th and 84th percentiles of the data. }
    \label{fig:4_3_2}
\end{figure*}

We separated the data by their total mass and \rsfr, or local sSFR. For total stellar mass, we set the boundary values as $10^{9.75}~M_{\odot}$, $10^{10.2}~M_{\odot}$, and $10^{10.7}~M_{\odot}$, which roughly separated all the spaxels into four equal parts. For sSFR and \rsfr, the binning was based on the 25\%, 50\%, and 75\% of the spaxels. 
Fig. \ref{fig:4_2_dmet} shows the potential secondary parameters versus the residual of rMZR. We use a sliding box of 0.05 dex to derive the median rMZR curve for most \msfd values, discarding the smallest and largest 0.5\% in \msfd since there are few data points there. Residuals of rMZR are then obtained relative to this median curve. The shaded cyan bands stand for the standard deviation of the data. In the top panel, the total stellar mass shows a positive correlation with the residual metallicity. Less massive galaxies tend to have lower metallicities with the same \msfd, and even the upper bounds of their standard deviations are below zero. This is consistent with previous works. In the middle and bottom panels, however, local sSFR and \rsfr do not show significant correlations with the residual metallicities. For both of them, although the cyan curves are slightly lower than zero at higher x-values, they are generally flat and very close to zero. They imply the non-existence of a resolved FMR: local sSFR and \rsfr do not drive systematic changes in metallicity. We also plot the median curves of sSFR and \rsfr versus metallicity residuals in different total mass bins. Surprisingly, with both parameters, only the dashed yellow curves, which represent the median curves of the lowest total mass bin, show slight deviation from the zero line. For the other 3 bins with higher total mass, the curves are even flatter. Table \ref{table:4_2} presents the partial correlation coefficients between these physical properties and the residual of local metallicity ($\Delta \rm \, log(O/H)$), controlling for \msfd. The partial correlation coefficient is defined as 

\begin{equation}
   \rho_{ZX|Y} = \frac{\rho_{ZX}-\rho_{ZY}\rho_{XY}}{\sqrt{1-\rho_{ZY}^2}\sqrt{1-\rho_{XY}^2}},
\end{equation}
where the $\rho_{ZX}$ is the Spearman's rank correlation coefficient between Z and X \citep{spearman1904proof}. The partial correlation coefficient between total mass and metallicity residual is 0.69, indicating a strong residual dependence on total stellar mass. For sSFR and \rsfr, the absolute values of partial correlation coefficients are both less than 0.2, suggesting that there could be weak correlations between metallicity and SFR at fixed \msfd. Considering different total mass bins, we find that the partial correlation coefficients of higher mass bins are all lower than 0.1. Explaining less than 1\% of the variance of the rMZR, sSFR, and \rsfr are practically uncorrelated to the metallicity residuals at $M_*\gtrsim 10^{9.75}~M_{\odot}$.

To explore potential tertiary dependencies beyond the secondary dependence on total mass, we further plot the rMZR across different total mass bins. In Fig. \ref{fig:4_2_3}, the gray 2D histograms present the distribution of spaxels in a certain total mass bin. The cyan solid curves show their median values, derived using the same method as in Fig. \ref{fig:4_2_dmet}. 
In the upper panels, the spaxels are further separated by \rsfr, and the dashed colored lines show the median of different \rsfr sub-bins. In the low-mass bin, there is a trend that the median metallicity decreases when the \rsfr increases with fixed \msfd. In the other three bins, this trend disappears, as all the dashed lines are close to the cyan lines. In the lower panels, sSFR also shows a minor influence on metallicity, except when the galaxy's total mass is low, as shown in the lower-left panel. In that panel, the relation between sSFR and metallicity at fixed \msfd is more significant in the high \msfd range. 

This result is expected, as it is consistent with previous studies showing that stellar mass exerts a stronger influence on gas-phase metallicity than either the global SFR of the galaxy or \rsfr \citep{2022MNRAS.514.2298B,2024ApJ...977..175L}. Using a self-consistent metallicity calibration across our sample, we find no significant correlation between metallicity and either \rsfr or sSFR once $\rm M_*$ is controlled for. Only at the low mass end of our sample, there are weak correlations with \rsfr and sSFR. However, we do not consider the weak correlations as the existence of resolved FMR. First, correlations are quantitatively weak. Even in the lowest mass bin, the partial correlation coefficients are -0.05. This can only explain less than 0.3\% of the variance ($|\rho|^2<0.003$). Second, in Figs. 7 and 8, we do not find evidence supporting global FMR even in low mass bins. 

This helps reconcile the long-standing debate over whether the FMR exists locally. Different metallicity calibrations can yield different correlations (positive, negative, or no correlation) between SFR and metallicity \citep{2020ApJ...897...61T}. This calibration dependence complicates the interpretation of the FMR and may partly account for the different conclusions reported in the literature. While some studies reported a weak secondary dependence on \rsfr, the correlation diminishes when selecting the self-consistent metallicity calibrator. Also, these findings imply a physical picture in which the local metallicity of a resolved region is governed by long-term chemical evolution, rather than by short-timescale star-formation episodes traced by \ha over the last 10 Myr \citep{2024MNRAS.52710788B}. \rsfr can serve as a tracer of gas inflow and outflow, so the results favor the explanation that metal dilution and enrichment average out over a 10 Myr timescale, especially in massive systems with deeper potential wells and higher metal retention efficiencies \citep{2007ApJ...658..941D,2018ApJ...852...74B}. In galaxies with lower masses, higher \rsfr represents the prominent effects of pristine gas inflow and metal-rich gas outflow, causing lower metallicity in general. Our results thus support the scenario that long-term accumulation of heavy elements dominates the chemical abundance.

\begin{figure*}
   \centering
	\includegraphics[width=\textwidth]{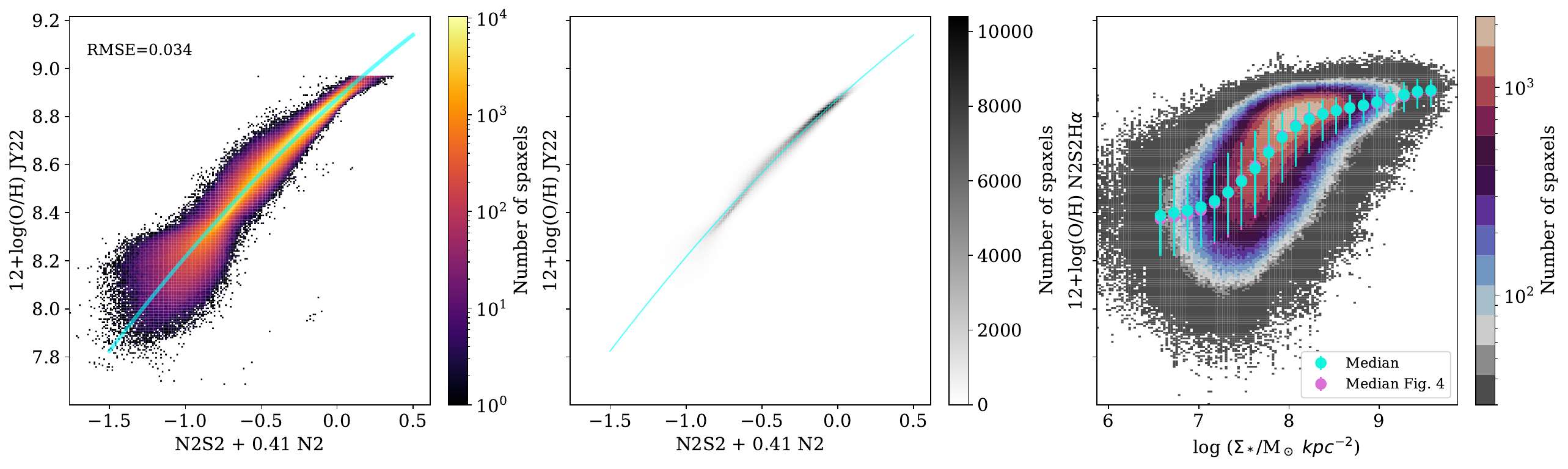}  
    \caption{\textit{Left: }$\rm R = log(\hbox{[N II]}\lambda\ 6584/\hbox{[S II]}\lambda\lambda\ 6716,6731)+0.41~log(\hbox{[N II]}\lambda\ 6584/H\alpha\,\lambda\ 6563)$ versus \citetalias{ji2022correlation} metallicities. The cyan line is the best-fit quadratic curve for all the spaxels. Data are color-coded by their number density. The root mean square deviation in metallicity is 0.034. \textit{Middle:} Same as the left panel, but use the linear scale to show the density of points. \textit{Right: }The rMZR derived using the new calibration. The cyan points with vertical bars show the median values and scatters of the 0.15 dex bins. The underlying magenta points are the same as the points in Fig. \ref{fig:4_1}, representing the rMZR derived using JY22 metallicity. The cyan points and the magenta points are very similar to each other.}
    \label{fig:5_1} 
\end{figure*}

\subsection{Integrated fundamental metallicity relation}

In galaxies, local metallicity is constrained by both local and global physical properties. The \rsfr-Z$_{local}$ relation can be extended to integrated SFR-Z$_{galaxy}$ relation \citep{2019ApJ...878L...6S}. Also, some of the previous studies show that the integrated MZR does not correlate with (s)SFR at high mass end \citep{2008ApJ...672L.107E,2013MNRAS.432.1217P}, which is similar to our results in last section.
In this section, we investigate the integrated FMR of MaNGA star-forming galaxies. 

To select star-forming galaxies, we keep those with more than 50 spaxels that meet our selection criteria in Sec. \ref{sec:2_3}. This selects 4548 galaxies among the MaNGA sample. We use the metallicity at the R$_{eff}$ of each galaxy as its integrated metallicity. To avoid the bias caused by random sampling, we selected spaxels between 0.75 R$_{eff}$ and 1.25 R$_{eff}$ and took their average to represent the metallicity at R$_{eff}$. Using the median will not change the result of this study. Fig. \ref{fig:4_3_1} illustrates the integrated scaling relations. The integrated MZR is shown in the left panel, with each dot color-coded by its total SFR. Although the low-mass end tends to have deeper color, this is more of a reflection of the star formation main sequence (SFMS, \citealt{2016ApJ...821L..26C}) rather than an SFR dependence. In the right panel, we plot the SFR versus metallicity, and show the median metallicities as functions of SFR in 10 M$_*$ bins. As the sample size is limited, we use solid lines to represent medians derived from more than 50 data points, and dashed lines to represent medians derived from fewer than 50 data points. Evidently, the SFR is independent of metallicity for high-mass galaxies ($M_*>10^{10.2}~M_{\odot}$), where the median curves are almost flat. For lower-mass galaxies, the curves are generally flat but with some wiggles, and the median values tend to be high at low SFR. However, the lowest SFR ends of most of the curves are derived from a few data points. Besides, even accounting for the bumpiness of the median curves, the median curves of each mass bin differ so much that they barely overlap in metallicity. For instance, all median metallicities of the lowest-mass bin are lower than any median metallicities in the second-lowest-mass bin. Basically, any secondary dependence on SFR is very weak, if it exists at all. 

Similar to Fig. \ref{fig:4_2_dmet}, in Fig. \ref{fig:4_3_2} we plot the integrated SFR and sSFR versus the residual of the integrated MZR. The median curves of the residual are shown in a solid magenta line. The shaded regions above and below the lines illustrate the 16th and 84th percentiles of the data. While the scatters decrease with increasing SFR, the medians remain at zero, showing that the residual metallicity is totally independent of the SFR. For sSFR, the median values also stick to zero and only go slightly negative at the highest sSFR end, which could also be due to the small sample size there. While it is consistent with most of the studies using IFU \citep{sanchez2012califa}, the result disfavors the existence of an FMR. We also plot the median curves from different total mass bins, using yellow, orange, red, and black dashed lines. Unlike \cite{2017ApJ...844...80B}, we do not observe a significant deviation from zero when the SFR or sSFR is low. In our resolved study shown previously, the rMZR slightly depends on \rsfr when galaxies' total masses are low. However, in Fig. \ref{fig:4_3_2}, the SFR within R$_{eff}$ does not influence the integrated MZR, for both low-mass and high-mass galaxies. 

One potential reason for this observational result is that the timescale of chemical enrichment for some elements may be very different from that of \ha-traced SFR. And the fluctuation of SFR may not be driven by the inflow of metal-poor gas. 
Previous studies concluding the existence of an FMR could be affected by observational bias of single-fiber spectroscopy, as most of the star-forming galaxies exhibit higher metallicities at their centers \citep{2014A&A...563A..49S,2017MNRAS.469..151B}. Similarly, the SFR in the center of galaxies could not reflect the total SFR of galaxies \citep{2013A&A...554A..58S,2014ApJ...797..126S,2016ApJ...827...35T}. Moreover, different metallicity calibrations could lead to different results for the dependence on SFR \citep{kewley2008metallicity,2019MNRAS.484.3042S}.  Using the \citetalias{ji2022correlation} metallicity calibration, we find that there is no secondary dependence on SFR for MZR, both locally and globally.



\section{A new metallicity proxy}
\label{sec:5}

In Section \ref{sec:3}, we found that the \nsh metallicity measurement provides a good match to both the direct method and the \citetalias{ji2022correlation} metallicity. Thus, it is useful to construct a new metallicity proxy similar to the one based on \nsh, as it is easier to apply than Bayesian inference. 
We provide a new metallicity proxy using N2 and N2S2. N2S2 has been used as a metallicity indicator in several previous works, both in empirical calibration \citep{2006A&A...459...85N} and theoretical calibration \citep{kewley2002using,2007MNRAS.381.1719V,2009MNRAS.398..949P}. This line ratio is insensitive to dust, offering a good indicator, as the dust attenuation on 1-2 kpc scales as observed by MaNGA-like observations is difficult to explain using a single extinction curve \citep{ji2023need}. Although \cite{kewley2002using} suggests that N2S2 has dependence on ionization parameter, \cite{2006A&A...459...85N} shows that the ionization parameter dependence can be ignored. N2 is also a commonly-used metallicity tracer \citep{2004MNRAS.348L..59P,marino2013o3n2,curti2017new,brazzini2024metallicity}. In our fitting, we parameterize the contribution between two line ratios: log(\NII/\SIId) and log(\NII/\Halpha) by a coefficient in front of the latter, and then parameterize metallicity as a quadratic function of this linear combination of N2S2 and N2 line ratios. We emphasize that this yields a single calibration rather than a blend of two independent calibrators: the relative weight between the two line ratios and the quadratic coefficients are all fitted simultaneously, so that 12+log(O/H) is a smooth, single-valued function of one combined line ratio, R:
\begin{equation}
    \label{eqs:r}
    \begin{split}
    \rm R = &\,\rm log(\hbox{[N II]}\lambda\ 6584/\hbox{[S II]}\lambda\lambda\ 6716,6731)\\
    &+0.410\,\rm log(\hbox{[N II]}\lambda\ 6584/H\alpha\,\lambda\ 6563)
    \end{split}
\end{equation}

The left and middle panels of Fig. \ref{fig:5_1} plot all the spaxels in the plane of \citetalias{ji2022correlation} metallicity vs. this line ratio combination, color-coded by their densities of distribution. The best-fitting equation is 

\begin{equation}
   \label{eqs:n2s2}
   \begin{split}
   \rm 12+log(O/H)=-0.0790R^2 
    + 0.574R + 8.87.
   \end{split}
\end{equation}
The formal error on the coefficients are all negligible compared to the root-mean-square scatter around the median relation ($\sim$ 0.034 dex). Unlike previous studies on N2S2 that use cubic or quartic functions, a quadratic function already provides a good fit over the full metallicity range probed, with a root mean square deviation of 0.034, and the relation passes through the densest (lightest-coloured) regions across the whole range. 

We note that the scatter of the relation increases slightly towards the lowest metallicities. This most likely reflects the intrinsically reduced sensitivity of N2S2 to metallicity in this regime: at high metallicity nitrogen is dominated by secondary production while sulfur is a primary element, making their ratio sensitive to metallicity, whereas at low metallicity both are dominated by primary production and the ratio varies only weakly with O/H \citep{kewley2002using,2006A&A...459...85N}. This is a property of the line ratio itself and does not bias the median relation used below; the inclusion of N2 further helps to tighten the calibration in this regime. 

In the right panel of Fig. \ref{fig:5_1}, we utilize this new empirical calibration and the \msfd to plot the rMZR. The cyan points shows the median values of the new calibration in a sliding box, in the same way as in Fig. \ref{fig:4_1}. The magenta points from Fig. \ref{fig:4_1} show a good consistency with the new median values down to the lowest-metallicity end, meaning that this empirical relation, combining \ha, \sii, and \nii, can serve as a more available substitute for \citetalias{ji2022correlation} calibration.


\section{Conclusions}
\label{sec:6}
We study the resolved and integrated stellar mass - gas-phase metallicity relations (MZR) and their secondary dependence for $\sim$ 4550 galaxies and $\sim$ $\rm 3.5\times10^6$ spaxels from the SDSS-IV/MaNGA survey, using the photoionization model-based metallicity calibration derived by \cite{ji2022correlation}.

\begin{enumerate}
   \item The photoionization model-based metallicity calibration derived by \cite{ji2022correlation} produces self-consistent results when using different combinations of optical line ratios. It shows the most consistency with the direct method, compared to several commonly used empirical calibrations.  
   \item We presented the spatially-resolved MZR in Fig. \ref{fig:4_1} and fitted it using three different functional forms as listed in Table \ref{table:4_1}. We confirmed that the different rMZRs reported in previous works mostly originate from different metallicity calibrations used.
   \item We found a strong correlation between the total stellar mass and the residuals of rMZR. However, there is no significant correlation between resolved sSFR or \rsfr and the residuals of rMZR, especially in galaxies whose total mass is larger than $10^{9.75}~M_\odot$. In galaxies whose total mass is smaller than $10^{9.75}~M_\odot$, a slight anti-correlation is presented between the \rsfr (or sSFR) and the metallicity residual. We conclude that the resolved FMR is not significant using the \citetalias{ji2022correlation} metallicity calibration, and the previous findings of the existence of the resolved FMR might be due to the calibrations used.
   \item We constructed the integrated MZR and checked for the existence of any secondary dependence on SFR for selected MaNGA star-forming galaxies. We find that the integrated data show no evidence for a statistically significant secondary dependence on SFR. The metallicity residuals of the integrated MZR do not show any correlation with the (s)SFR, even after controlling for the total mass.
   \item A new empirical calibration is built using a quadratic function of the combination of N2S2 and N2. It provides a good fit in intermediate and high metallicity regimes (12+log(O/H) > 8.4), with the RMSE of 0.034 for all the spaxels. It could serve as a simplified substitution of the \citetalias{ji2022correlation} method for deriving metallicity.
   \end{enumerate}

Combining the resolved and integrated results, the stellar mass, SFR, and gas-phase metallicity depict a picture of long-term chemical evolution. Our results suggest that metallicity is determined by the accumulated production of heavy elements, while the gas inflow and outflow, traced by SFR, reach an equilibrium so that there are no significant trends between SFR and metallicity. 
The self-consistency across multiple line ratio combinations provides validation for our adopted photoionization model, and strengthens confidence in the resulting rMZR and its lack of secondary dependence on SFR.

\section*{Data availability}

The photoionization model grids and the corresponding \textsc{Cloudy} input files used in this work are publicly available on Zenodo at \url{10.5281/zenodo.21717332}. The MaNGA data analyzed in this work are publicly available as part of SDSS Data Release 17.

\begin{acknowledgements}
We thank the anonymous referee for providing a detailed and insightful report.
We acknowledge the grant support by the National Natural Science Foundation of China (NSFC; grant No. 12373008, 12425302), the support by the Research Grant Council of Hong Kong (Project No. 14302522, 14303123), by the Direct grant from the Faculty of Science of CUHK, and the support by the Hong Kong Jockey Club Charities Trust through the project, JC STEM Lab of Astronomical Instrumentation and Jockey Club Spectroscopy Survey System. RY acknowledges support by the Hong Kong Global STEM Scholar Scheme (GSP028). Z.S.L. acknowledges the support from Hong Kong Innovation and Technology Fund through the Research Talent Hub program (PiH/022/22GS). Y.C. is supported by the Direct Grant for Research (C0010-4053720) from the Faculty of Science, the Chinese University of Hong Kong. 

Funding for the Sloan Digital Sky 
Survey IV has been provided by the 
Alfred P. Sloan Foundation, the U.S. 
Department of Energy Office of 
Science, and the Participating 
Institutions. 

SDSS-IV acknowledges support and 
resources from the Center for High 
Performance Computing  at the 
University of Utah. The SDSS 
website is www.sdss4.org.

SDSS-IV is managed by the 
Astrophysical Research Consortium 
for the Participating Institutions 
of the SDSS Collaboration including 
the Brazilian Participation Group, 
the Carnegie Institution for Science, 
Carnegie Mellon University, Center for 
Astrophysics | Harvard \& 
Smithsonian, the Chilean Participation 
Group, the French Participation Group, 
Instituto de Astrof\'isica de 
Canarias, The Johns Hopkins 
University, Kavli Institute for the 
Physics and Mathematics of the 
Universe (IPMU) / University of 
Tokyo, the Korean Participation Group, 
Lawrence Berkeley National Laboratory, 
Leibniz Institut f\"ur Astrophysik 
Potsdam (AIP),  Max-Planck-Institut 
f\"ur Astronomie (MPIA Heidelberg), 
Max-Planck-Institut f\"ur 
Astrophysik (MPA Garching), 
Max-Planck-Institut f\"ur 
Extraterrestrische Physik (MPE), 
National Astronomical Observatories of 
China, New Mexico State University, 
New York University, University of 
Notre Dame, Observat\'ario 
Nacional / MCTI, The Ohio State 
University, Pennsylvania State 
University, Shanghai 
Astronomical Observatory, United 
Kingdom Participation Group, 
Universidad Nacional Aut\'onoma 
de M\'exico, University of Arizona, 
University of Colorado Boulder, 
University of Oxford, University of 
Portsmouth, University of Utah, 
University of Virginia, University 
of Washington, University of 
Wisconsin, Vanderbilt University, 
and Yale University.\\

\textit{Softwares:} Starburst99 \citep{leitherer1999starburst99}, \textsc{Cloudy} \citep{ferland20172017}, Astropy \citep{Astropy_Collaboration_and_Price-Whelan_The_Astropy_Project_2022}, Numpy \citep{2020NumPy-Array}, Matplotlib \citep{Hunter_Matplotlib_A_2D_2007}, 
Scipy \citep{Scipy}, Pingouin \citep{Pingouin}
, extinction \citep{barbary_extinction_2016}.
\end{acknowledgements}

\bibliographystyle{aa} 
\bibliography{all_ref}

\end{document}